\documentclass[]{interact}

\usepackage{epstopdf}
\usepackage[caption=false]{subfig}

\usepackage[numbers,sort&compress]{natbib}
\usepackage{float}
\bibpunct[, ]{[}{]}{,}{n}{,}{,}
\renewcommand\bibfont{\fontsize{10}{12}\selectfont}
\makeatletter
\def\NAT@def@citea{\def\@citea{\NAT@separator}}
\makeatother

\theoremstyle{plain}

\theoremstyle{definition}

\theoremstyle{remark}

\begin{document}

\articletype{ARTICLE TEMPLATE}

\title{Real-time Path Planning of Driver-less Mining Trains with Time-dependent Physical Constraints}

\author{
\name{Xiaojiang Ren\textsuperscript{a*}\thanks{CONTACT Changxin Gao Email: xjren@xidian.edu.cn}, 
Hui Guo\textsuperscript{a},
Sheng Kai\textsuperscript{a}
and Guoqiang Mao\textsuperscript{a}}
\affil{
\textsuperscript{a}Guangzhou Institute of Technology, Xidian University, Guangzhou, China; 
\textsuperscript{b}School of Artificial Intelligence and Automation, Huazhong University of Science and Technology, Wuhan, China}
}

\maketitle

\begin{abstract}
While the increased automation levels of production and operation equipment have led to improved productivity of mining activity in open pit mines,
the capacity of mine transport system become a bottleneck.
The optimisation of mine transport system is of great practical significance to reduce the production and operation cost 
and improve the production and organizational efficiency of mines.
In this paper we first formulate a multi-objective optimisation problem for mine railway scheduling by introducing a set of mathematical constraints.
As the problem is NP-hard, we then devise a Mixed Integer Programming based solution to solve this problem, and develop an online framework accordingly.
We finally conduct test cases to evaluate the performance of the proposed solution. Experimental results demonstrate that the proposed solution is efficient and able to generate train schedule in a real-time manner.
\end{abstract}

\begin{keywords}
mine railway; train scheduling; mixed integer programming; intelligent transportation system
\end{keywords}

\section{Introduction}

The optimisation of railroad transportation system in open pit mines is an important part of the optimisation of complex large system in open pit mines.
It is of great practical significance to reduce the production and operation cost, 
together with improving the production and organizational efficiency of mines.
In recent years,
while the increased automation levels of production and operation equipment have led to improved productivity,
part of the mine transport system bottleneck problem began to highlight,
especially on the basis of the static road network analysis method of the traditional routing optimisation planning models and algorithms.
When faced with more complex and variable scheduling decisions,
traditional solutions are easily limited,
and difficult to get global realistic optimal results.

Various research work, 
such as mixed integer programming and dynamic programming, 
have been conducted to solve urban railroad transportation optimisation problems in the literature
~\cite{GH2020, HZN2021, WYZ2021,XFC2020,YAW2019, YT2016}.
Compared with urban railway networks,
mine railway networks have limited amount of tracks, 
most of which are bi-directional single-track. 
This leads to some particular problems as follows.
\begin{itemize}
    \item Track allocation: Giving most tracks are bi-directional single-track, the transportation capacity is limited by the number of receiving and departure tracks, the number of siding/meetpoints, and the length of single-tracks. Moreover, the capacity is also restricted by track maintenance.
    \item Train conflict: For a bidirectional single-track, there are mainly two conflicts between trains. First, a pre-defined time difference is necessary to ensure safe operation when trains running in the same direction; Second, trains in opposite directions cannot be in the same segment at any time.
\end{itemize}

Given the problems stated above, the mine railroad scheduling problem is a complex optimisation problem,
and the results under different decisions are far from each other. 
With purpose of less running time of all trains and maximizing the total amount of rolling stock,
this paper proposes an integrated model to optimize train timetable and track allocation.
The main contributions of this paper are listed as follows. 

\begin{itemize}
    \item We first formulate a multi-objective optimisation problem for mine railway scheduling by introducing a set of mathematical constraints.
    \item As the problem is NP-hard, we then devise a MIP-based solution to solve this problem in a real-time manner.
    \item We finally conduct test cases to demonstrate the validity and effectiveness of the solution.
\end{itemize}

In the following sections, 
Section~\ref{relatedwork} presents a review on the railway scheduling and train timetabling problems. 
Section~\ref{problem} introduces the system model and describes the mining railroad optimisation problem. 
Section~\ref{mathmodel} provides the mathematical formulation for the objective and the set of constraints.
Section~\ref{solution} presents a description of all the modules that constitute the developed solution, in order for all the readers to get an understanding of the entire process.
Section~\ref{evaluation} presents a summary of the result for the tests carried out to evaluate the proposed solution.
Section~\ref{conclusion} is dedicated to the conclusions.



\section{Related Works}\label{relatedwork}

An enormous number of studies have been conducted on railway scheduling and train timetabling problems~\cite{CFK2016, Harrod2011, KWSZ2015,KM2017,LMB2019, QLG2018}.
The research work can be classified into three main categories: 
the train scheduling and rescheduling problem,
the periodic and non-periodic timetabling problem,
and the passenger train and freight train timetabling problem~\cite{LLX2020}.

Train scheduling is an offline problem that determines the arrival and departure times for trains at each station before the schedule is executed, e.g., ~\cite{BES2021,LLY2020,NZG2015,SDC2018, WDY2018}. 
With a planned schedule, rescheduling is a real-time problem that aims to determine detailed train movements and timetables to minimize train deviations, e.g.,~\cite{KWS2015, WSS2021, XXZ2020,ZG2019}.
Sanat et al.~\cite{SDC2018}  studied a train scheduling problem in a large national railway network, and 
presented two flexible heuristics based on a Mixed Integer Program formulation for local optimisation to improve infrastructure utilization.
Wang et al.~\cite{WSS2021} proposed a train rescheduling optimisation model in the case of the vehicle breakdown on a metro line. Efficient rescheduling strategies including flexible short-turning and adding backup trains are particularly formulated into the model. 
Bersani et al.~\cite{BES2021} formalized demand-oriented scheduling and rescheduling models in order to propose a dynamic timetable, and proposed a min-max method to address operational constraints related to train capacity, train speed limits, train transfers, possible conflict in the track section use, with the main objective to minimize the travel time.

Periodic timetabling requires that most or all train paths repeat in time with a certain period (e.g., 12 hours), e.g.,~\cite{HKL2021,PBD2019,SG2017,SS2020,ZSZ2013}. However, as it becomes difficult to obtain effective periodic schedules when dealing with interruptions or conflicts (e.g. track maintenance), 
a non-periodic timetable becomes more appropriate, e.g.,~\cite{CT2012,Harrod2012}.
Huang et al.~\cite{HKL2021} integrated stop planning, service planning, and scheduling in a periodic timetabling problem and modelled it as a Mixed Integer Linear Programming formulation to minimize the average travel delay of passengers. They then developed a genetic algorithm supported by a scheduling heuristic to solve the problem for better scalability and efficiency. 

Because railways provide both passenger and freight services, there are naturally passenger train and freight train timetabling problems, e.g.,~\cite{BKS2016,BWG2020,KZS2016,MD2011,OP2018,SF2012}. 
Mu and Dessouky~\cite{MD2011} introduced two mathematical formulations to cope with the rapidly increasing freight demand for railway transportation, and presented several heuristics that can significantly reduce the solution time of the exact method carried out by CPLEX yet produce a satisfactory solution quality.
Bešinović et al.~\cite{BWG2020} introduced the integrated passenger and freight train timetable adjustment problem which handles both passenger as well as freight trains,
and developed a mixed integer linear programming model to simultaneously retime, reroute and cancel trains in the network.

These research lines are viewed from different perspectives but are not truly independent of each other. 
For example, passenger train scheduling is usually a periodic timetabling problem.
Li et al.~\cite{LLX2020} studied a non-periodic freight train scheduling problems, in which a schedule is planned for freight trains and can be different in different periods of the day. 
The proposed method considers car flow transfer between consecutive trains and the shipment delivery time requirement. A tabu search algorithm is developed that extends the applicability of the proposed optimisation method for large-scale problems. Experiments on real-world instances of the Menghua railway indicate the effectiveness of the proposed algorithm.

\section{Problem Description}\label{problem}

\begin{figure}[H]
\includegraphics[scale=0.70]{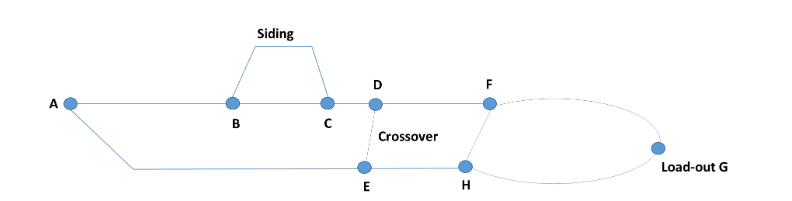}
\caption{Railway Network Sample}
\label{fig:systemmodel}
\end{figure}   
As shown in fig.~\ref{fig:systemmodel}, 
it is a sample layout of mining railway network, 
where trains start from station A, travel via the tracks, load mine from load-out G, and return back to station A.

The mining railroad optimisation problem is to determine the best feasible timetable for a set of trains in order to load mines as many as possible. 
The related constraints are to satisfy restrictive operational constraints (e.g., track capacity, travel speed, safe distance etc.) and to avoid possible conflicts when using each single track.

\subsection{System Model}

Define a physical network $G^*=(S,W)$ with a set of nodes S and a set of links W. 
The set S consists of switch stations ($S_u \subset S$), and load-outs ($S_l \subset S$) in physical network. 
$W=\{(i,j)\}$ is the set of links in physical network, 
where $(i,j)$ stands for link connecting node $i\in S$ to node $j\in S$.
The set W consists of mainline tracks ($W_u \subset W$), siding tracks ($W_w \subset W$) 
and crossovers ($W_o \subset W$) in physical network.
For each link $(i,j)$, let 
\begin{itemize}
    \item $d_{ij}$ be the capacity of $(i, j)$: Here capacity is the maximum number of trains that can stand on the link $(i, j)$ at any point of time.
    \item $f (i , j)$ be the travel time from i to j, where f is the speed profile based function.
    \item $f (j, i)$ be the travel time from j to i, where f is the speed profile based function.
\end{itemize}
In order to model the business operation of mine loading, for each load-out $i \in S_l$, let $P_i$ be the average loading time for a train.

Let $Q = \{0, 1, \cdots, |Q|-1\}$ be the set of time instants, where time is discretized into discrete time instants of length g minutes. For instance if we take g = 5 then a period of 1 hour would be represented by discrete set $Q = \{0, 1, 2, \cdots, 12\}$ in our model.

\subsection{Train Model}
Let $M = \{1, 2, \cdots, m\}$ be the set of real trains travelling in the physical network $G^*$.
For each train $m \in M$,we will get the following set of inputs:
\begin{itemize}
    \item $loSeq^m$: Sequence of scheduled load-outs to visit. 
    \item $depTime^m$: Scheduled departure time. 
    \item $depLoc^m$: Scheduled departure location. 
\end{itemize}

As we have a cyclical network and train changes direction (turns back) after going to a load-out, in order to model this behaviour we will break down the train journey into different parts. Each part will be represented by a different model train. Thus, if a train $m$ goes to $n$ load-outs, its whole journey will be represented by a set of $n+1$ model trains called $T_m$, where each model train represents a specific segment of train $m$’s journey. 
Accordingly, one business rule will be added that any model train can depart only after its predecessor has finished its journey.

For example, considering the case of a train $m$ starting from station A and going to two load-outs G1 and G2, its journey will be modelled as follows:
\begin{table}[H] 
\caption{Model Train Example \label{tab:train}}
\begin{tabular}{cccc}
\toprule
\textbf{Model Train}	& \textbf{Departure Node} & \textbf{Destination Node} & \textbf{Predecessor Train}\\
\midrule
\(t_{m1}\)		& A	& G1		& null\\
\(t_{m2}\)		& G1	& G2		& \(t_{m1}\)\\
\(t_{m3}\)		& G2	& A		& \(t_{m2}\)\\
\bottomrule
\end{tabular}
\end{table}

From now on, we will use train to refer to model train only, unless specified otherwise. 
Thus, for each train $t \in T$, we will have the following set of information: 
\begin{itemize}
    \item $depQ_t$: Scheduled departure time. 
    \item $dep_t$: Scheduled departure location node. 
    \item $dest_t$: Final destination node of the train, after which it will disappear from the network.
    \item $t_{pre}$: Predecessor train of train t.
\end{itemize}

\subsection{Time Space Network}\label{timespace}

We here consider a time-space network $G = (N, A)$, where N denotes the node set and A denotes the arc set. For each node $i \in N$, let $Outb_i \subseteq A$ be the set of all the outbound arcs of node $i$, and $Inb_i \subseteq A$ be the set of all the inbound arcs of node $i$.

Given a physical network $G^*$, we then can construct the time-space network $G$ as follows:
\begin{itemize}
    \item For each load-out in $s \in S_l$, we replace it with two separate node $s_{in}$ (entry node) and $s_{out}$ (exit node), and add a load-out link ($s_{in}$, $s_{out}$) into W. Let $W_l \subset W$ be the set of load-out links. By splitting the entry and exit point nodes for each load-out, we ensure that trains are staying at load-out for the required loading time. At the same time, we replace the related inbound links $(i, s) \in W$ to $(i, s_{in})$ and the related outbound links $(s, j) \in W$ to $(s_{out}, j)$. Accordingly, we have $f(i, s_{in})=f(i, s)$, $f(j, s_{out})=f(j, s)$, and $f(s_{in}, s_{out})=P_s$, where $P_s$ indicates the required loading time.
    
    \item For each link $(i, j) \in W$ whose capacity $d_{ij} >1$, we break down this link into smaller links by adding $\lfloor d_{ij}-1 \rfloor$ dummy nodes into S and replacing the link $(i, j)$ with $\lfloor d_{ij}-1 \rfloor$ dummy links (with proportional travel time) as well. Note: After this operation, the capacity of each arc is 1. This operation guarantees that on every arc at any point of time only one train travels on that arc, which will in turn guarantee that trains maintain a safe headway separation while travelling in the same direction.
    
    \item For each siding track $(i, j) \in W_w$, we add a siding node $i'$ to $S$ and replace the siding track $(i, j)$ with two separate arcs $(i, i')$ and $(i', j)$. Let $S_w \subset S$ be the set of siding nodes. Accordingly, we have $f(i, i')=f(i', j)=f(i, j)/2$ and $f(j, i')=f(i', i)=f(j, i)/2$. For each link $(i, j) \in W$, let $Iden(i, j)$ be the set of identical arcs in $A$.
    
    \item For each node $s \in S$, we add $|Q|$ corresponding nodes $\{s^0, ..., s^{|Q|-1}\}$ in $N$. 
    For each train $t \in T$, we add corresponding virtual source node $s_0^t$ in $N_s \in N$, where $s_0^t$ represents the source node where train $t$ departs. We also add one sink $s_1$ into $N$, where $s_1$ represents the sink node where every train terminates its journey.
    
    \item For each link $(i, j) \in W$, we add following transit arcs into $Iden(i, j) \subset A$:
     $$(i^k, j^{k+f(i,j)}),\ for\ all\ k = 0, ..., |Q|-1\ and\ k+f(i,j) \leq |Q|-1$$
     $$(j^k, i^{k+f(j,i)}),\ for\ all\ k = 0, ... , |Q|-1\ and\ k+f(j,i) \leq |Q|-1$$
     
     \item For each link $(i, i^{'}) \in W_w$, we then update $Iden(i, i^{'})$ as follows:
     $$Iden( i, i^{'} )= Iden( i^{'}, j )=Iden( i, i^{'} ) \cup Iden(i^{'}, j )$$
     
     \item For each siding node $i \in N$, we add following waiting arcs into $A_w \subset A$. In this way, we allow trains to wait/dwell on sidings.
     $$(i^k, i^{k+1}),\ for\ all\ k = 0, ..., |Q|-2$$
     
     \item For any load-out $s \in S_l$ with loop capacity $cap_s$, we add $(cap_s -1)$ waiting arcs between $(s_{in}^k, s_{out}^k)$ , for all $k = 0, ..., |Q-2|$. In this way, we allow trains to wait/dwell in the loops.
     
     \item For each station node $i \in S_u$, we add an arc from that node to the sink, signifying that allows train cancelling in case of deadlock
     (i.e. we can cancel a train with a very high penalty).
     Thus we add these train disappearing arcs into $A_d \subset A$:
     $$(i^k, s_1 ),\ for\ all\ k = 0, ..., |Q|-1$$
     
     \item For each train $t \in T$, we add an arc from the source to the scheduled starting node of that train, and we add these starting arcs into $A_s \subset A$:
     $$(s_0^t, j^k ),\ for\ all\ k \geq depQ_t,\ and\ j\ is\ the\ scheduled\ starting\ node(i.e.,\ j=dep_t)$$
     
     \item In all the above cases, whenever an arc $(i, j) \in A$ is added to the time-space network, we also add this arc to the Outbound arc set of node $i (i.e. Outb_i)$ and to the Inbound arc set of node $j (i.e. Inb_j)$.
\end{itemize}

\section{Mathematical Model}\label{mathmodel}

According to the definitions made in the previous section, the objective and the set of constraints are now presented in a formal manner.

\subsection{Attributes}
For each $(i^k , j^l) \in A$ and a given train $t \in T$, there exists a cost attribute $c^t_{i^kj^l} $ as follows:
$$
c^t_{i^kj^l}= \left\{\begin{matrix} 
\gamma*(l-k), \forall t \in T, \forall (i^k, j^l) \in A-A_w-A_d-A_s \\
\alpha *( |Q| -\lfloor k/(60/g)\rfloor),\forall t \in T, \forall (i^k, i^{k+1}) \in A_w \\
\rho * time_{left},\ if\ dest_t =i \ or\ k=Q,\forall t \in T, \forall (i^k, s_1) \in A_d\\
M,\ if\ dest_t \neq i, \forall t \in T, \forall (i^k, s_1) \in A_d\\
\beta *(l- depQ_t),\forall t \in T, \forall (s_0^t, j^l) \in A_s
\end{matrix}\right.
$$
where $(i^k , j^l)$ denotes an arc in the time space network representing possible movement (of a train) starting from node $i$ at time $k$ and terminating at node $j$ at time $l$.
The cost $\alpha *( |Q| -\lfloor k/(60/g)\rfloor)$ implies that we give weighted penalties for waiting/dwelling, where $\alpha$ is constant parameter.
The cost $\beta *(l- depQ_t)$ implies the penalty for late departure, where $\beta$ is constant parameter.
The cost $\rho * time_{left}$ implies that we try to route trains as close as to the destination.
The cost $\gamma*(l-k)$ implies that we try to route trains along the shortest-time path.

\subsection{Variables}
For each $ (i^k, j^l)\in A $ and a given train $t\in T$, define the following binary decision variable:
$$
x^t_{i^kj^l}= \left\{\begin{matrix} 
1,if\ train\ t\ will\ travel\ on\ arc(i^k,j^l)\ during\ k\ to\ l \\
0,otherwise
\end{matrix}\right.
$$

\subsection{Objective}
Given the above cost attributes and variables, we then define the objective as follows:
$$
Min\sum_{t\in T}\sum_{(i^k, j^l)\in A}c^t_{i^kj^l}*x^t_{i^kj^l}
$$
\subsection{Constraints}
In particular, we formulate the related constraints as follows:
\begin{itemize}
    \item Flow conservation constraint: These constraints algebraically state that the sum of the flow through arcs directed toward a node plus that node's supply, if any, equals the sum of the flow through arcs directed away from that node plus that node's demand, if any.
    $$\sum_{(s^t_0,j)\in Outb_{s^t_0}}x^t_{s_0^tj} =1,\forall t \in T$$
    $$\sum_{(i,s_1)\in Inb_{s_1}}x^t_{is_1} =1,\forall t \in T$$
    $$\sum_{(i,j)\in Outb_i}x^t_{ij}-\sum_{(j,i)\in Inb_i}x^t_{ji}=0,\forall i \in N-N_s-\{s_1\},\forall t\in T$$
    
    \item Node capacity constraint: This constraint ensures that for each node excluding $s_1$, trains can be held safely without any collision or deadlock at any time point.
    $$\sum_{t \in T}\sum_{(j,i)\in Inb_i}x^t_{ji}\leq 1,\forall i \in N-\{s_1\}$$
    \item Arc capacity constraint: This constraint implies that for any track, load-out in the network, at any time instant $q \in Q$, there can only be at max one train travelling or staying.
    $$\sum_{t\in T}\sum_{(i^{'k},j^{'l})\in Iden(i,j)\&\{k\leq q-1, l\geq q\}}x^t_{i^{'k}j^{'l}}\leq 1, \forall q\in Q, \forall (i,j)\in W$$
    \item Train departure constraint: This constraint implies that any train $t$ with predecessor not null, which represents the return trip of a real train,
    has to depart at the time when its predecessor train completes its journey.
    $$x^t_{s^t_0dep^q_t}-x^{t_{pre}}_{dest^q_ts_1}=0,\forall q\in [depQ_t,Q], \forall t \in T \& t_{pre} \neq null$$
\end{itemize}

\section{Solution}\label{solution}

Given an offline scenario, i.e. optimise the train schedule for a pre-defined time window,
a traditional approach is to solve the above MIP problem directly. 
However, the above approach can not apply into the online scenario,
since the solving time often turn out to be unacceptable. e.g., it may take several days to generate a train schedule for the next 24-hour. 
Instead, we here propose a solution framework for the online scenario.
\begin{figure}[H]
\includegraphics[scale=0.65]{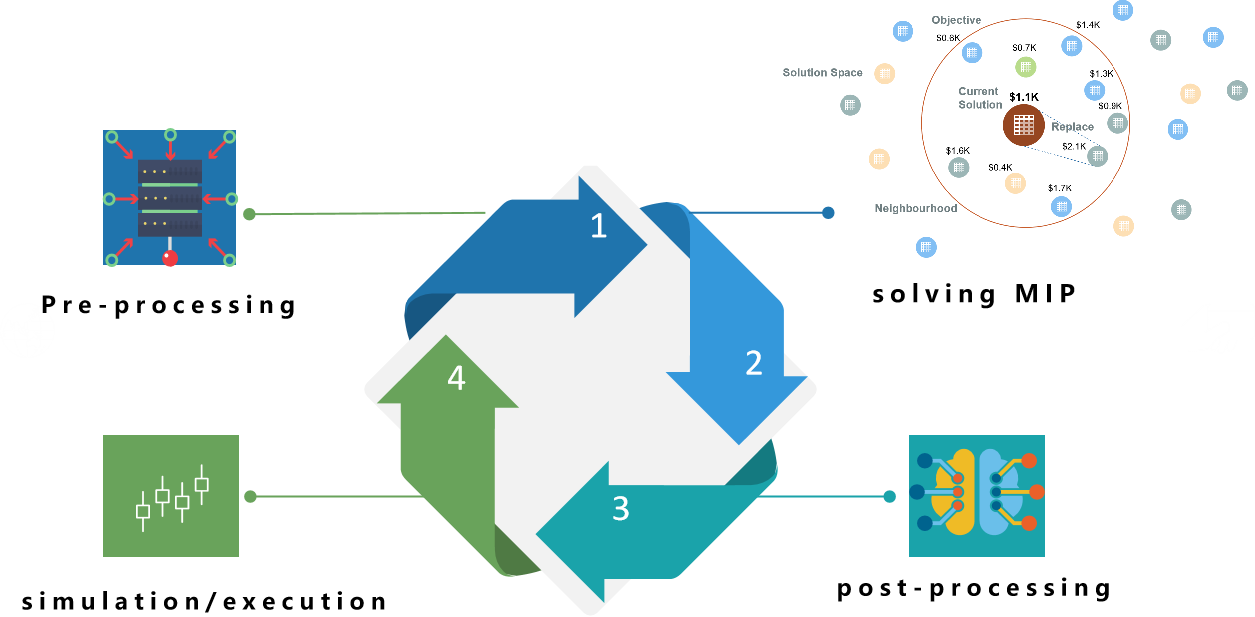}
\caption{Solution Framework}
\label{fig:framework}
\end{figure}  
As shown in fig.~\ref{fig:framework}, it runs iteratively. 
The cycle length is pre-defined (e.g. every $5$ minutes), which is decided as per the business requirements (e.g. problem size, algorithm running time, etc). 
For each cycle, it contains $4$ steps as follows.

\subsection{Pre-processing}
\begin{figure}[H]
\centering
\includegraphics[scale=0.9]{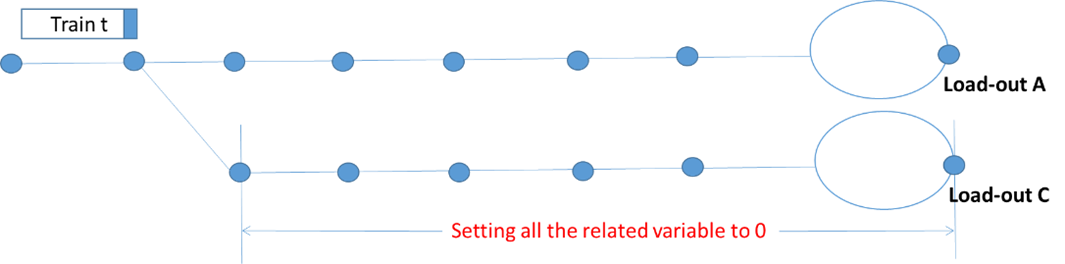}
\caption{Preprocessing Sample 1}
\label{fig:pre1}
\end{figure}  
\begin{figure}[H]
\centering
\includegraphics[scale=0.9]{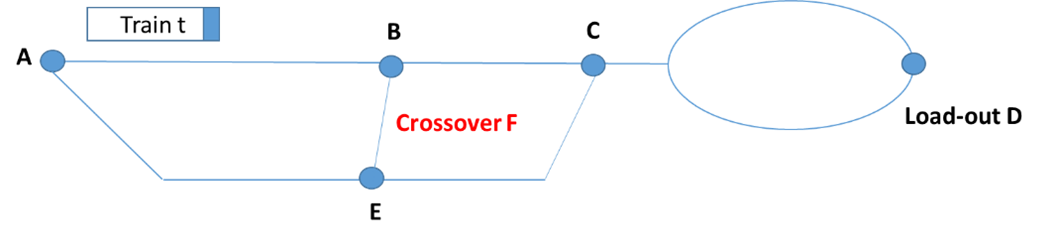}
\caption{Preprocessing Sample 2}
\label{fig:pre2}
\end{figure}  
In the pre-processing step, we use the current status as input,
and formulate the above
MIP problem, which complexity changes exponentially when variable size varies. 
To further reducing the running time, we adopt the following approaches to reduce the variables:
\begin{itemize}
    \item Initializing variables only for arcs on which train will travel: for each train, we only consider the reasonable arcs according to its destination. For example,
    in fig.~\ref{fig:pre1} the train $t$ is heading for load-out A. We then can set all the unreasonable variables (i.e., those variables associated to the arcs heading for load-out C) to 0.
    \item Removing variables corresponding to unrealistic movements: consider the train $t$ in fig.~\ref{fig:pre2} going towards load-out, which is not allowed to move on track F.
    We then fix variables corresponding to such invalid movements to 0.
\end{itemize}

\subsection{Solving MIP}
In this step, we solve the formulated MIP problem via mathematical optimisation solver (e.g. Gurobi, Xpress, Cplex, etc). 
To better utilize the solver, we also integrate more strategies as follows: 
\begin{itemize}
    \item Warm start:  supply hints to help solver find an initial solution, which consist of pairs of variables and values, known as a warm start. The hints may come from soft business rules or human experience.
    \item Solver tuning: use offline data to tune the solver, and adopt the solver setting for the online running.
\end{itemize}

\subsection{Post-processing}
After getting the MIP results, 
we then need to translate the math-style results into trains' schedule solution. In particular, we need to combine trains with the related predecessor trains to get complete movements. 
As shown in table~\ref{tab:train}, we need to combine the results of $t_{m1}$, $t_{m2}$, $t_{m2}$ 
to get the schedule of train $m$. 

\subsection{Simulation/execution}
For each given solution, 
we then validate the schedule further via simulation, 
and add more details including maintenance allocation.
We finally trigger an actual execution by communicating the detailed schedule to trains.

\section{Experiments and Results}\label{evaluation}
In this section, we evaluate the performance of the proposed solution.
\subsection{Experimental Environment Setting}
\begin{figure}[H]
\centering
\includegraphics[scale=0.6]{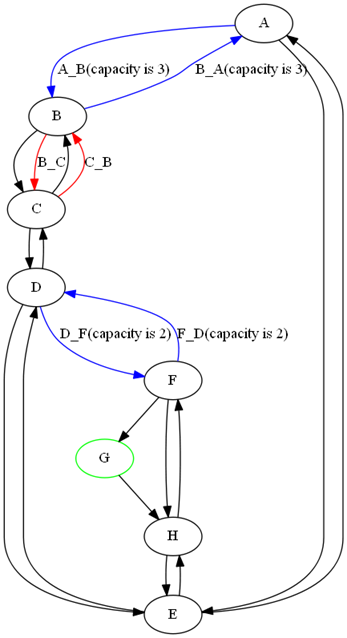}
\caption{Physical Network with Capacity}
\label{fig:capacitynetwork}
\end{figure}

\begin{figure}[H]
\centering
\includegraphics[scale=1]{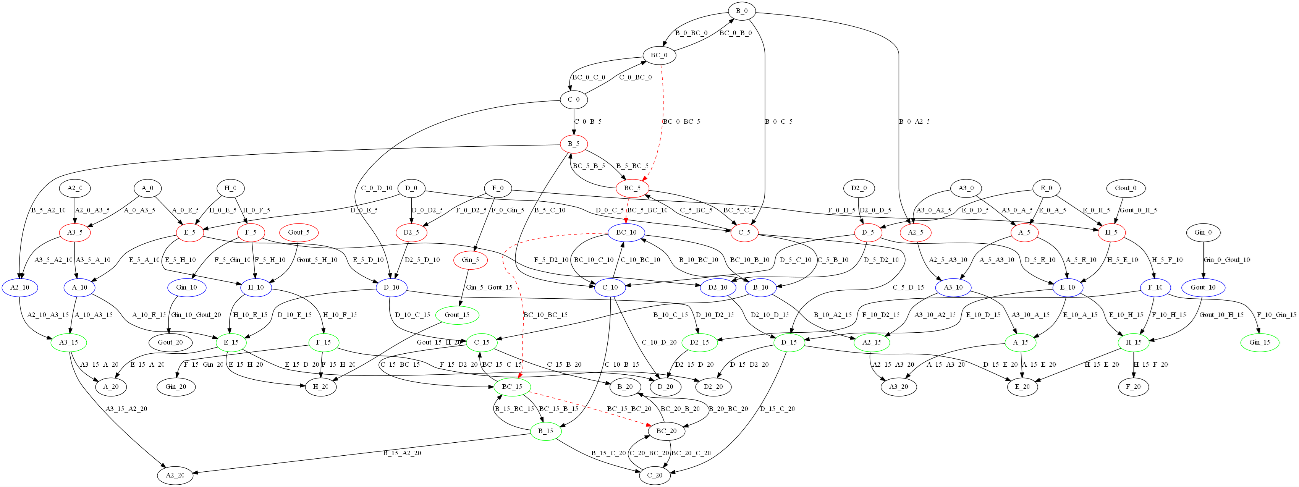}
\caption{Constructed Time-space Network}
\label{fig:timespacenetwork}
\end{figure}

We here consider the sample network in fig.~\ref{fig:systemmodel}, 
and has track capacity shown in fig.~\ref{fig:capacitynetwork}. 
As per the subsection~\ref{timespace}, we then construct the time-space network of 20-minute time window in fig.~\ref{fig:timespacenetwork}, for which the time instant length is 5 minutes.

We then evaluate the proposed solution via the following test cases. 
From case 1 to case 4, we vary trains to schedule, generate the train schedule, 
and demonstrate the solution in time-space network.
\subsection{Test Case 1}
\begin{figure}[H]
\includegraphics[scale=0.9]{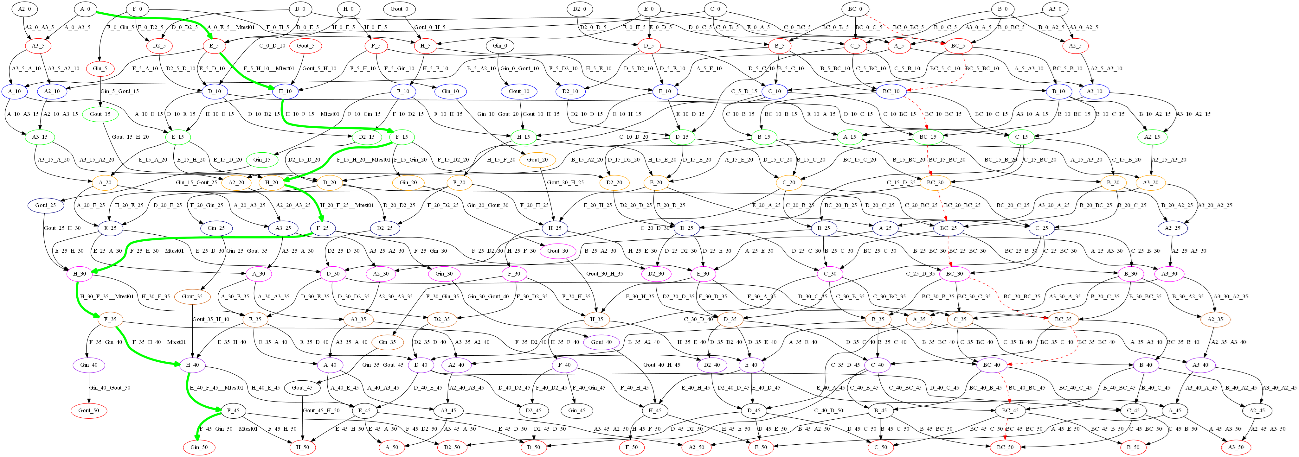}
\caption{Case 1: Train Schedule}
\label{fig:test1}
\end{figure}

\begin{table}[H] 
\caption{Case 1: Train Info\label{tab:test1}}
\begin{tabular}{cccc}
\toprule
\textbf{Train Name}	& \textbf{Departure Node} & \textbf{Departure Time} & \textbf{Destination Node}\\
\midrule
\(Mtest01\)		& A	& 0		& G\\
\bottomrule
\end{tabular}
\end{table}
Given a single train info in table~\ref{tab:test1}, we are able to get the solution in less than one minute, where the train schedule for train Mtest01 is demonstrated via green-path in fig.~\ref{fig:test1}.

\subsection{Test Case 2}
\begin{figure}[H]
\includegraphics[scale=0.9]{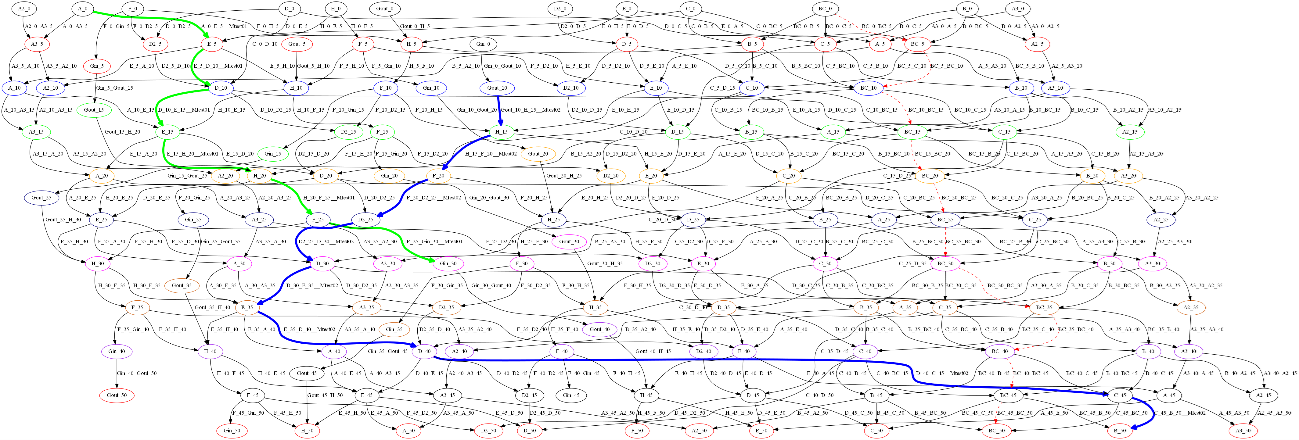}
\caption{Case 2: Train Schedule}
\label{fig:test2}
\end{figure}

\begin{table}[H]
\caption{Case 2: Train Info\label{tab:test2}}
\begin{tabular}{cccc}
\toprule
\textbf{Train Name}	& \textbf{Departure Node} & \textbf{Departure Time} & \textbf{Destination Node}\\
\midrule
\(Mtest01\)		& A	& 0		& G\\
\(Mtest02\)		& G	& 0		& B\\
\bottomrule
\end{tabular}
\end{table}
We have two trains to schedule in table~\ref{tab:test2}, where train Mtest01 goes towards load-out G and train Mtest02 returns from load-out G. 
We are able to get the solution in less than one minute, where the train schedule for train Mtest01/Mtest02 are demonstrated via green-path/blue-path respectively in fig.~\ref{fig:test2}.

\subsection{Test Case 3}
\begin{figure}[H]
\includegraphics[scale=0.9]{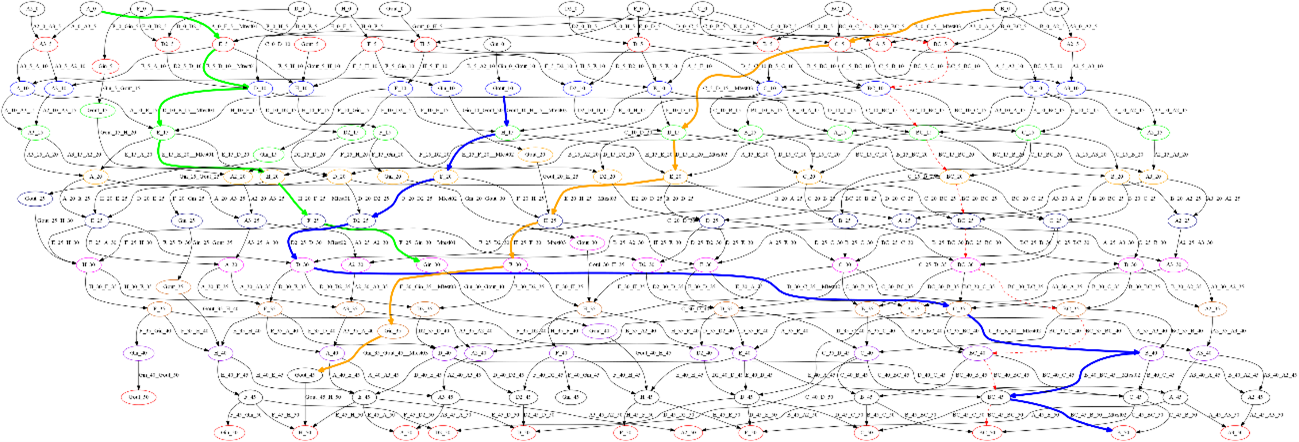}
\caption{Case 3: Train Schedule}
\label{fig:test3}
\end{figure}

\begin{table}[H]
\caption{Case 3: Train Info\label{tab:test3}}
\begin{tabular}{cccc}
\toprule
\textbf{Train Name}	& \textbf{Departure Node} & \textbf{Departure Time} & \textbf{Destination Node}\\
\midrule
\(Mtest01\)		& A	& 0		& G\\
\(Mtest02\)		& G	& 10		& B\\
\(Mtest03\)		& B	& 0		& G\\
\bottomrule
\end{tabular}
\end{table}
We have three trains to schedule in table~\ref{tab:test3}: both train Mtest01 and Mtest03 go towards load-out G from different start station; train Mtest02 goes back from load-out G with departure time constraint. 
We are able to get the solution in less than one minute, and the train schedule are demonstrated via in fig.~\ref{fig:test3}. In particular, train Mtest01/Mtest02/Mtest03 schedule are demonstrated via green-path/blue-path/orange-path respectively.

\subsection{Test Case 4}
\begin{figure}[H]
\includegraphics[scale=0.9]{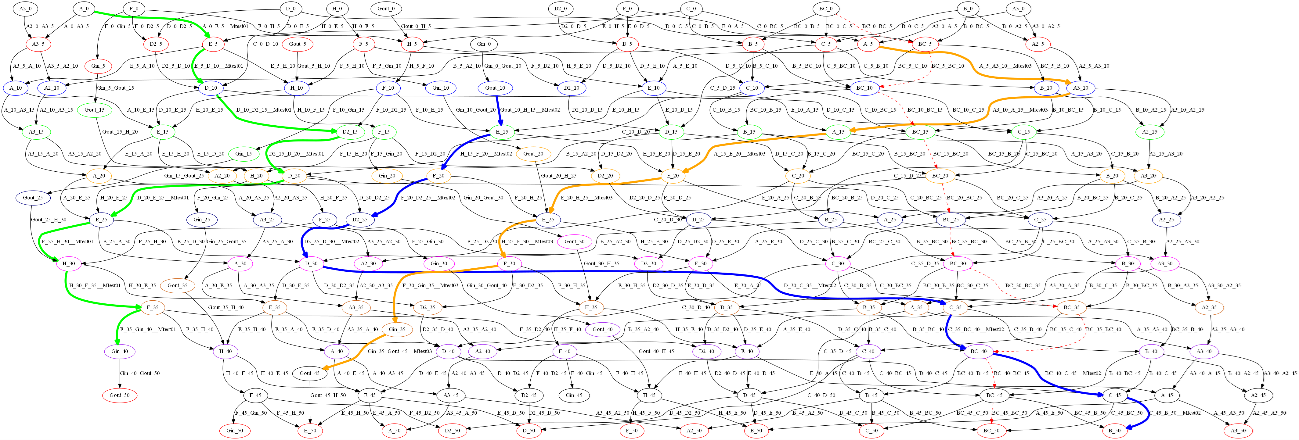}
\caption{Case 4: Train Schedule}
\label{fig:test4}
\end{figure}

\begin{table}[H] 
\caption{Case 4: Train Info\label{tab:test4}}
\begin{tabular}{cccc}
\toprule
\textbf{Train Name}	& \textbf{Departure Node} & \textbf{Departure Time} & \textbf{Destination Node}\\
\midrule
\(Mtest01\)		& A	& 0		& G\\
\(Mtest02\)		& G	& 10		& B\\
\(Mtest02\)		& A	& 0		& G\\
\bottomrule
\end{tabular}
\end{table}
Given three trains info in table~\ref{tab:test4}, 
both train Mtest01 and Mtest03 go towards load-out G with same start station; train Mtest02 goes back from load-out G with departure time constraint.
Similar to case 3, 
we are able to get the solution in less than one minute, 
and demonstrate the train schedule in fig.~\ref{fig:test4},
where train Mtest01/Mtest02/Mtest03 schedule are showed via green-path/blue-path/orange-path respectively.

\section{Conclusions}\label{conclusion}
This work studied an online mining railway optimisation problem. 
We first formulated the problem as a novel MIP problem, and proposed an online solution accordingly.
The solution has been evaluated via test cases.
However, it is worthy mentioning that our model is an initial integrated optimisation model for mine railway scheduling, and station track allocation planning, 
which has more generalisation space for some specific purposes. 
Further research will focus on the following several aspects:
\begin{itemize}
    \item consider more practical constraints, such as ensuring first-come-first-serve policy near load-outs.
    \item develop an effective heuristic algorithm to increase the efficiency of the solution.
\end{itemize}



\section*{Disclosure statement}
No potential conflict of interest was reported by the author(s).


\section*{Funding}

This work was supported by the National Natural Science Foundation of China under Grant U21A20446; Guangzhou Applied Basic Research Program under Grant 202201011786.

\end{document}